\documentclass[9pt, conference]{IEEEtran}
\IEEEoverridecommandlockouts

\usepackage{cite}
\usepackage{amsmath,amssymb,amsfonts}
\usepackage{algorithmic}
\usepackage{graphicx}
\usepackage{textcomp}
\usepackage{xcolor}
\usepackage{multirow}
\usepackage{hyperref}
\usepackage{fancyhdr}
\def\BibTeX{{\rm B\kern-.05em{\sc i\kern-.025em b}\kern-.08em
    T\kern-.1667em\lower.7ex\hbox{E}\kern-.125emX}}

\fancypagestyle{firstpagefooter}{
    \fancyhf{} 
    \setlength{\footskip}{40pt} 

    \fancyfoot[R]{%
        \parbox{\dimexpr\columnwidth\relax}{%
            \rule{\dimexpr0.7\linewidth}{0.05pt}\\ 
            The implementation and pre-trained weights on public datasets are available at 
            \href{https://github.com/khanld/chunkformer}{\textit{github.com/khanld/chunkformer}}.
        }%
    }
}

\begin{document}

\title{ChunkFormer: Masked Chunking Conformer For Long-Form Speech Transcription\\
}
\author{\IEEEauthorblockN{Khanh Le\IEEEauthorrefmark{1}, Tuan Vu Ho, Dung Tran\IEEEauthorrefmark{2} and Duc Thanh Chau\IEEEauthorrefmark{3}\IEEEauthorrefmark{4}}
\IEEEauthorblockA{\IEEEauthorrefmark{1}ZaloAI, Vietnam}
\IEEEauthorblockA{\IEEEauthorrefmark{2}Independent Researcher}
\IEEEauthorblockA{\IEEEauthorrefmark{3}Ho Chi Minh City University of Science \\}
\IEEEauthorblockA{\IEEEauthorrefmark{4}Vietnam National University, Ho Chi Minh City
}
\texttt{\{khanhld218, tuanvu.ksvt92, dung.n.tran.xyz\}@gmail.com, ctduc@fit.hcmus.edu.vn} \\

\textit{Corresponding author: Duc Thanh Chau}

}


\maketitle

\begin{abstract}
Deploying ASR models at an industrial scale poses significant challenges in hardware resource management, especially for long-form transcription tasks where audio may last for hours. Large Conformer models, despite their capabilities, are limited to processing only 15 minutes of audio on an 80GB GPU. Furthermore, variable input lengths worsen inefficiencies, as standard batching leads to excessive padding, increasing resource consumption and execution time. To address this, we introduce ChunkFormer, an efficient ASR model that uses chunk-wise processing with relative right context, enabling long audio transcriptions on low-memory GPUs. ChunkFormer handles up to 16 hours of audio on an 80GB GPU, 1.5x longer than the current state-of-the-art FastConformer, while also boosting long-form transcription performance with up to 7.7\% absolute reduction on word error rate and maintaining accuracy on shorter tasks compared to Conformer. By eliminating the need for padding in standard batching, ChunkFormer's masked batching technique reduces execution time and memory usage by more than 3x in batch processing, substantially reducing costs for a wide range of ASR systems,  particularly regarding GPU resources for models serving in real-world applications.
\end{abstract}


\section{Introduction}
\label{sec:intro}
Recent advancements in scaling attention-based models have introduced challenges in memory consumption, execution time, and performance, particularly with long input sequences. For industrial cloud-based ASR systems, the cost of scaling to support millions of users can reach millions of dollars in GPU resources. Current advanced models struggle with long-form transcription, as their ability to process extended speech remains limited. This issue is exacerbated when dealing with highly variable input lengths, where padding in naive batching causes resource consumption and processing time to increase significantly. For instance, decoding a batch with 1-hour and 1-second audios is essentially equivalent to processing two 1-hour audios due to padding.
Conformer model \cite{49414}, which combines convolutional layers for local features and self-attention for global context, achieves impressive results in speech recognition. However, this success comes with significant computational and memory demands. The quadratic complexity associated with sequence length in the attention mechanism severely limits the maximum audio duration that the Conformer model can handle.
The Efficient-Conformer \cite{burchi2021efficient} refines the Conformer downsample module by achieving 8x subsampling. This extended subsampling is achieved by replacing the original convolution module with a stride-2 depthwise convolution module in the last two encoder stages. They further employed grouped attention to alleviate the computational burden of early attention layers on lengthy sequences, allowing them to reduce attention complexity from $\mathcal{O}(L^2d)$ to $\mathcal{O}(L^2d/g)$ for sequence length $L$, hidden dimension $d$, and group size parameter $g$.
\begin{figure}[ht]
    \vspace{-20pt}

  \centering
  \includegraphics[width=\linewidth]{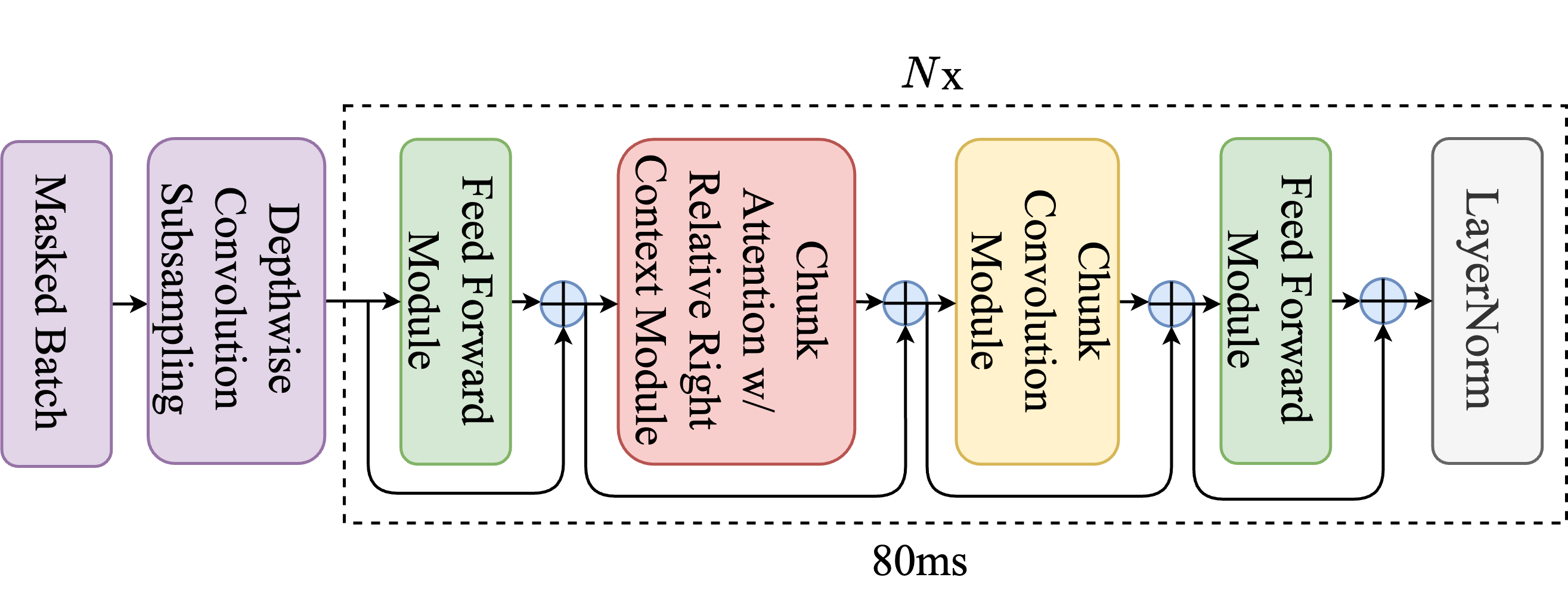}
\vspace{-20pt}
  \caption{The ChunkFormer model architecture}
  \label{fig:figure1}
    \vspace{-15pt}
\end{figure}
FastConformer \cite{rekesh2023fast} also increases the down-sampling rate to 8x utilizing the depthwise convolution but right at the start of the model. Additionally, this method decreases the quadratic complexity of the attention layer by integrating the attention mechanism from LongFormer \cite{beltagy2020longformer}, reducing the complexity for input sequences of length $L$ to $\mathcal{O}(L \times (128 \times 2 + 1) \times d)$, where 128 is the window size surrounding each token.  By switching to limited context attention, they extend the maximum duration that the model can process at once by 45x. 
Other efforts to transcribe long-form audio involve segmenting the audio into smaller parts using either fixed length or Voice Activity Detection (VAD), then sorting and batching audio files of similar lengths before model inference. For fixed-length segmentation, a buffer is employed to maintain historical context between segments. Conversely, WhisperX \cite{bain2022whisperx} proposes merging short VAD segments to maximize contextual relevance, optimizing the balance between segmentation length and context preservation.
While effective, the aforementioned approaches have some limitations. While FastConformer extends the maximum duration consumption, transcribing long audio sequences on low-memory GPUs remains unfeasible, and token-wise processing is inefficient for parallel processing. WhisperX avoids this challenge by employing segmentation decoding. However, because segments are decoded independently, this results in the loss of context information between segments. Additionally, both methods suffer from padding in batch processing, reducing efficiency and increasing resource consumption.
\thispagestyle{firstpagefooter}

In this paper, we represent ChunkFormer, a novel speech recognition model for long audio sequences: 
\begin{enumerate}
    \item \textbf{Endless Decoding:} ChunkFormer inherits chunk-based with relative right context processing to extend Longformer's attention mechanism \cite{beltagy2020longformer}, thereby enabling continuous decoding. This allows the model to process longer and unlimited audio duration on a low-memory GPU while retaining context information between chunks.
    \item \textbf{Masked Batch Decoding:} A novel masked batching technique integrated with ChunkFormer, which removes the use of padding associated with high-variant input lengths in standard batching, thus enabling the batch transcription of 1-hour and 1-second audios to use the same resources as a single 1-hour and 1-second audio, rather than the resources required for two 1-hour audios due to padding.
\end{enumerate}
ChunkFormer has proven its effectiveness, surpassing state-of-the-art ASR models on small and large-scale datasets in performance, speed, and resource consumption, demonstrating its scalability in real-world systems.

\section{Methods}
\label{sec:methods}
\subsection{Endless Decoding in ChunkFormer}
 \begin{figure}[ht]
    \vspace{-8pt}
  \centering
  \includegraphics[width=\linewidth]{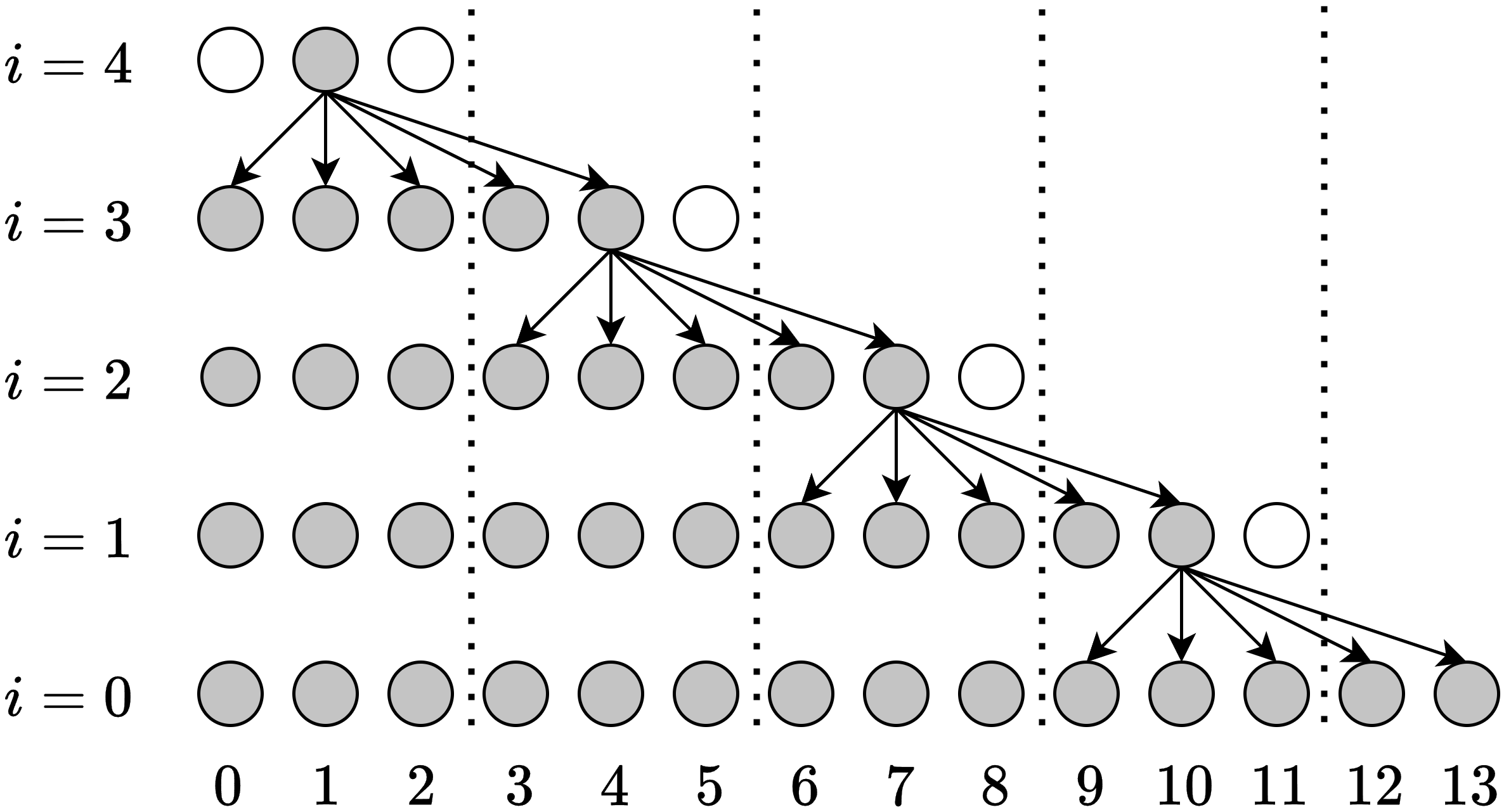}
\vspace{-13pt}

  \caption{Visualization of receptive field expansion in the $i^{th}$ encoder layer of ChunkFormer with $c = 3$, $r = 2$ and $N=4$.}
  \label{fig:rel_right_context}
\end{figure}
\noindent \textbf{Chunk Processing:} ChunkFormer is first inspired by the streaming approach for endless decoding. In the streaming process, input data arrives in chunks, and the conformer model generates outputs along with various model caches, such as attention cache, convolution cache, and input cache, to facilitate subsequent computations. This seemingly implies a sequential processing limitation due to the presence of caches. However, it's crucial to note that the caching mechanism in Conformer differs from that in sequential architectures like RNN or LSTM.
Let's assume we have a complete audio signal or more generally, a set of $n$ filter-bank chunk sequences $\{\mathbf{X}^{0}_{[0, c)}, \mathbf{X}^{1}_{[c, 2\times c)},..., \mathbf{X}^{n-1}_{[(n-1) \times c,  n \times c)}\}$, each of size $c$, arriving simultaneously. We can efficiently process these chunks in parallel along the batch axis rather than along the time axis. Assuming the maximum number of chunks that a GPU can handle at once is $m$, the input $\mathbf{X}$ to the model is: 
\begin{align}
\mathbf{X} &= \begin{bmatrix}\mathbf{X}^0 & \mathbf{X}^1 & \ldots & \mathbf{X}^{m-1}\end{bmatrix}
\end{align}
\textbf{ChunkFormer with relative right context:} Streaming functionality is achieved by applying chunk-based convolution \cite{Li2023DynamicCC} and attention mechanisms \cite{Chen2020DevelopingRS}. However, streaming models often underperform compared to full-context models due to the absence of future context. Increasing the chunk size can reduce this performance gap; however, \cite{le24_interspeech} demonstrated that frames at the end of each chunk still suffer from a lack of future context, resulting in suboptimal outcomes for these frames. We introduce the ChunkFormer attention mechanism, which integrates relative right context into the chunk-based system while preserving its endless decoding capability. Our attention mechanism differs from that in \cite{rekesh2023fast, beltagy2020longformer} in two key aspects. First, ChunkFormer employs overlapping context around each chunk instead of each token and eliminates the global token. These modifications are more efficient and computationally lighter, while still maintaining the endless decoding capability. Second, our right context attention cumulatively attends to subsequent chunks, unlike the static right context in the streaming system \cite{le24_interspeech,zeineldeen:chunked-aed}.
\begin{figure}[ht]
    \vspace{-15pt}

  \centering
  \includegraphics[width=\linewidth]{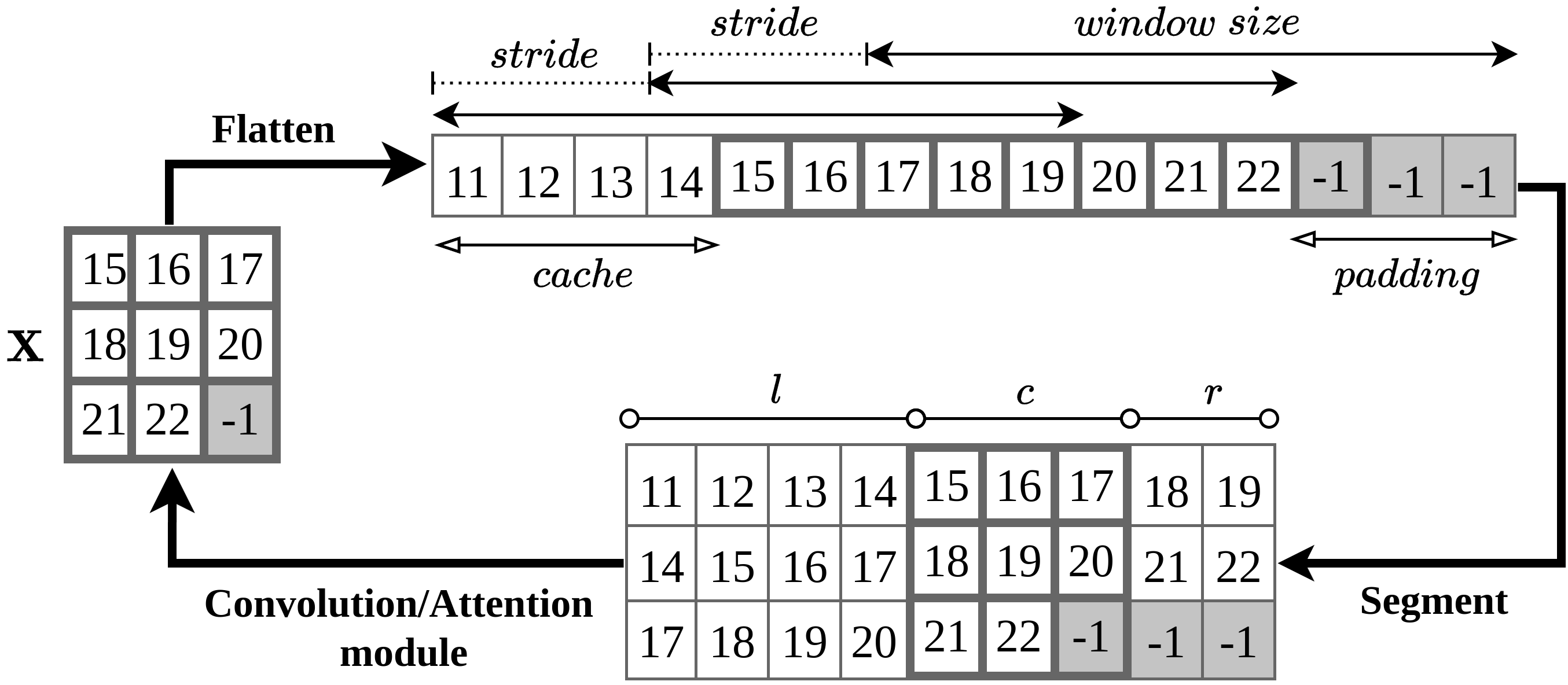}
    \vspace{-14pt}
  \caption{Illustration of Overlapping Chunk Transformation with \( l = 4 \), \( c = 3 \), and \( r = 2 \), assuming the first 15 frames have been decoded. Padding values of -1 are applied to the last chunk for chunk batching and to the flattened sequence for overlapping segmentation. Gray (padding) areas are masked during convolution and attention computation.
}
  \label{fig:OCT}
    \vspace{-12pt}
\end{figure}
Given an attention input chunk $\mathbf{X}^i = \begin{bmatrix}
x_{i \times c} & x_{i \times c + 1} & \dots & x_{i \times c + (c-1)}\end{bmatrix}$, the relative attention with right context computation for $x_j \in \mathbf{X}^i$ occurs as follows:
\begin{align}
e_{j,t} &= (\overbrace{x_{j} W_{q} W^{T}_k x^{T}_t}^{\text{self-attention}} + \overbrace{x_{j} W_{q} W^{T}_{\mathcal{R}} \mathcal{R}^{T}_{j-t}}^{\text{relative positional attention}} \nonumber \\ 
&+ \underbrace{u W^{T}_k  x^{T}_t + v W^{T}_{\mathcal{R}} \mathcal{R}^{T}_{j-t}}_{\text{bias term}})/\sqrt{d_k}
\end{align}

\begin{align}
\alpha_{j, t} &= \frac{\exp(\beta \times e_{j,t})}{\sum_{t=i \times c - l_{att}}^{(i+1) \times c + r - 1}\exp(\beta \times e_{j,t})}
\end{align}

\begin{align}
z_j &= \sum_{t=i \times c - l_{att}}^{(i+1) \times c + r - 1} \alpha_{j, t} W^{T}_v x_t^{T}
\end{align}
Here, $d_k$ denotes the feature dimension per head. We use weight matrices $W_q$, $W_k$, $W_v$, and $W_\mathcal{R}$ for queries, keys, values, and relative positional encodings, respectively. The model learns two additional parameters, $u, v \in \mathbb{R}^{d}$. The relative positional encodings are stored in a matrix $\mathcal{R} \in \mathbb{R}^{L_{\text{max}} \times d}$, where each row represents the encoding for a specific relative distance between two positions in the sequence.

To enable endless decoding with relative right context attention, we must compute the number of future frames accessible to each chunk. Otherwise, it will cause the mismatch with training to the output of the last chunks in $\mathbf{X}$ and to the rest of the subsequent chunks. For example, in Figure \ref{fig:rel_right_context}, with chunk size $c=3$, right context size $r = 2$, and the number of encoder layers $N=4$, we will need a total of 11 future frames to compute the output chunk of frames from 0 to 2. To compute the total future frames needed ($r_{rel}$), we use:
\begin{align}
r_{rel} &= r + max(c, r) \times  (N - 1)
\end{align}
Now, the input to the model with relative right context will be:
\begin{align}
\mathbf{X} = \begin{bmatrix} \mathbf{X}^0 & \mathbf{X}^{1} & \ldots & \mathbf{X}^{m-1} &  \mathbf{X}^{rel}_{[m \times c, m \times c + r_{rel})} \end{bmatrix}
\end{align}
and only the output of $\{{\mathbf{X}^0,  \dots, \mathbf{X}^{m-1}}\}$ is used to convert to text. \\ 
\textbf{Overlapping Chunk Transformation (OCT):} 
In ChunkFormer, all operations are sequentially independent, except for the depthwise convolution in the convolution module and the attention mechanism, which rely on sequential processing. The challenge lies in enabling interaction between chunks within the sequential operations, as each chunk in \( \mathbf{X} \) is treated as an independent sequence due to batch-axis merging. Addressing this, we introduce a transformation method applied directly to the input of depthwise convolution and the key-value pairs in the attention mechanism. This transformation follows the steps outlined in Figure \ref{fig:OCT}: the input is flattened, and overlapping segmentation is applied to generate inputs for sequential layers. The left context $l$ serves as model caches between chunks in a streaming manner, while the right context $r$ represents the relative right context described above. For the next decoding step \( \{\mathbf{X}^{m}, \dots, \mathbf{X}^{2m-1}\} \), the convolution and attention caches are derived directly from the index ranges \( [m \times c - l_{\text{conv}}, m \times c) \) and \( [m \times c - l_{\text{att}}, m \times c) \) of the flattened input, respectively, where convolution cache size $l_{\text{conv}} = \frac{\text{kernel size} - 1}{2}$ and $l_{\text{att}}$ is the attention cache size. Crucially, using caches from the previous decoding step avoids re-computation and matches the chunk-based training, eliminating the need for a historical context buffer. Thanks to OCT, ChunkFormer can process multiple chunks in parallel with the in-place creation of left/cache and right context, allowing it to decode audio input that exceeds the model's immediate consumption capacity in a seamless streaming manner. Importantly, by limiting context attention, ChunkFormer reduces the quadratic computation for a sequence of length $L$ to $\mathcal{O}(n \times c \times (l_{att}+c + r) \times d)$, where $L \approx n \times c$, thus enabling faster inference and increasing the maximum duration that the model can process at once.
 \begin{figure}[t]
       \vspace{-4pt}

  \centering
  \includegraphics[width=\linewidth]{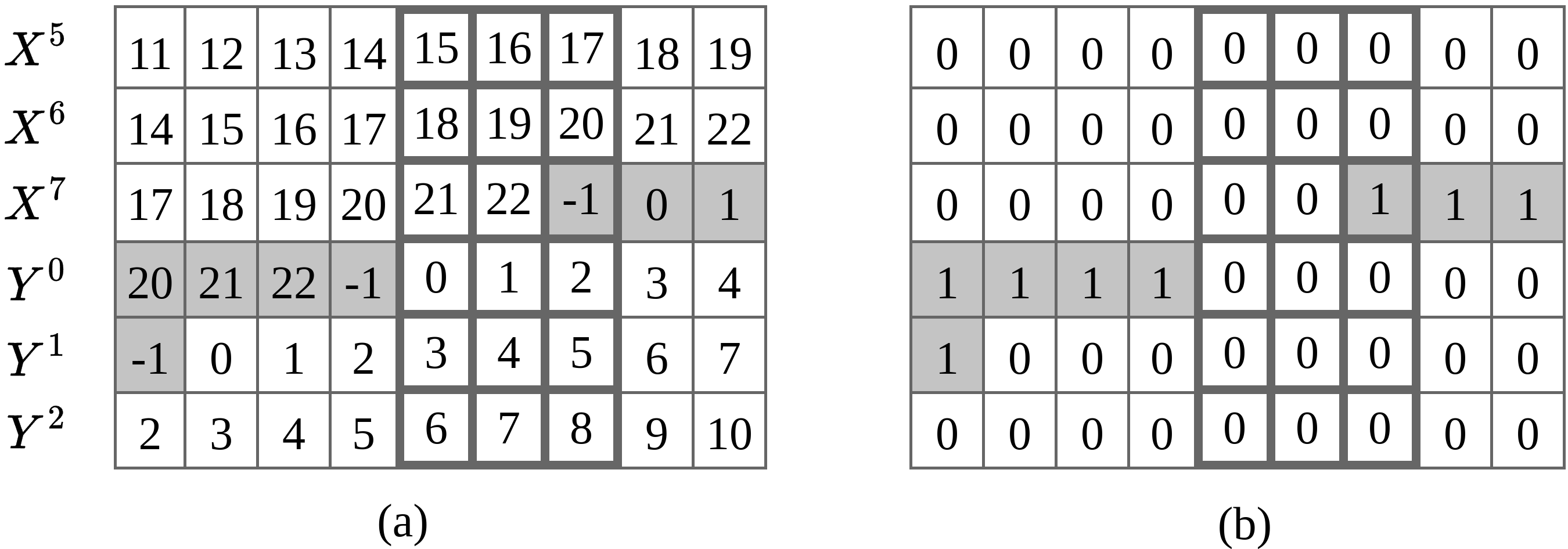}
      \vspace{-17pt}
  \caption{Masked batch with $l = 4$, $c = 3$, and $r = 2$. (a) shows the output from the OCT, and (b) displays the corresponding mask. Bold-bordered areas represent the current batch of chunks, with $m_x = 5$, $n_x = 8$, and $m_y = 3$. The numbers represent frame indices, and since $\mathbf{X}$ consists of 23 frames, a padding value of -1 is applied to $\mathbf{X}^7$ to complete a chunk. Gray areas are masked during the convolution and attention computations using (b) mask.}
  \label{fig:maskedchunk}
      \vspace{-12pt}

\end{figure} 
\subsection{Masked Batch Decoding}
\noindent \textbf{Chunk Batching: } Recall that ChunkFormer treats audio as a batch of equal-sized chunks. A batch of two varying-length audios $\mathbf{X}$ and $\mathbf{Y}$ can be represented as a batch of chunk sequences $\begin{bmatrix}  \mathbf{X}^{m_{x}} & \mathbf{X}^{m_{x} + 1} & \ldots &  \mathbf{X}^{n_x - 1} &  \mathbf{Y}^0 & \ldots  \mathbf{Y}^{m_y - 1}\end{bmatrix}$. This approach eliminates memory waste caused by padding in highly duration-imbalanced audio. We assume that the first $m_x$ chunks of $\mathbf{X}$ have already been processed, and the remainder of $\mathbf{X}$, along with the first $m_y$ chunks of $ \mathbf{Y}$, serves as input to the model. \\
\textbf{Mask out overlapping segments:} One issue with chunk batching is that the OCT segmentation process creates overlapping regions between consecutive audio segments at the start and end of the audio, as shown in Figure \ref{fig:maskedchunk}a. Consequently, this leads to incorrect left and right contexts in convolution and attention computations. While using a simple for-loop to perform OCT on each audio is a solution, it is slow, especially when applied to each encoder layer, and worsens as the number of audios increases. 
Therefore, we propose using a mask to zero out the overlapping regions from the segmentation output (Figure \ref{fig:maskedchunk}b). This mask is generated based on the audio length (e.g., 23 in $ \mathbf{X}^7$) for masking the right side and the number of active frame indices (white squares, e.g., 5 in $ \mathbf{Y}^1$ and 8 in $ \mathbf{Y}^2$) for masking the left side. Since the mask is created once and used across all encoder layers with batch processing, we achieve up to a 2.5x speedup in model inference time compared to the for-loop approach for a batch of 100 audios, each 10 seconds long, and $N = 17$.
\subsection{ChunkFormer Convolution}
ChunkFormer leverages FastConformer's subsampling and convolution modules updates, yet falls short of Conformer's performance in full context setting. To bridge this gap, we increased the kernel size of convolution modules to 15  and the filters of subsample module to encoder dimensions, effectively mimicking the Conformer convolution module's receptive field after downsampling. For better stability in long training of large models, we leverage layer normalization instead of batch normalization in the convolution modules.

Limited context attention extends audio processing length significantly, but large inputs to subsample convolution remain a memory bottleneck. Batch splitting into mini-batches is a solution but requires a batch size greater than one. FastConformer implementation tackles this by splitting along the channel axis and manually applying the depthwise spatial convolution sequentially. However, even splitting the batch and channel size to 1, these approaches still struggle with extremely long audio sequences on low-memory GPUs and lack the efficiency of framework-based matrix tensor processing. Masked batching solves this challenge perfectly since long audio is split into small chunks and processed independently in the subsample module. This allows batch splitting without worrying about memory limitations and leverages the inherent parallelism of frameworks, rather than the manual approach used in FastConformer.
\section{Experiments}
\label{sec:experiments}
\subsection{Experimental Setups}
\begin{table}[h]
\vspace{-10pt}
\centering
\caption{\label{tab:testset} Experimental Datasets. [SM] stands for short-medium form and [L] stands for large form.}
\vspace{-5pt}

    \begin{tabular}{ l  l  l }
    
     & \multicolumn{1}{l}{\textbf{Small scale}}  & \multicolumn{1}{l}{\textbf{Large scale internal}} \\ 
    \hline
    \textbf{Train} & 960h Librispeech (LS) & 25.000h Vietnamese \\
    \hline
    \multirow{5}{*}{\textbf{Test}} & [SM] 11h LS test set & [SM] 2.8h Reading \\
                                    & [L] 3h Tedlium-v3 & [SM] 11h Conversation\\
                                    & [L] 39h Earnings-21  & [SM] 7.5h Telephony\\
                                    & [L] 119h Earnings-22 & [L] 3.2h Youtube \\
    \hline
    \end{tabular}
\vspace{-4pt}

\end{table} 
\noindent \textbf{Datasets}: We evaluate ChunkFormer on small-scale and large-scale training datasets. While small-scale datasets are the well-known public, large-scale datasets are our internal collection, comprising extensive Vietnamese-labeled speech from various domains, including reading, conversation, telephony, and Youtube. The test sets are categorized into short-medium and long-form. Detailed specifications are provided in Table \ref{tab:testset}. \\
\textbf{Model:} We utilize the CTC-AED model following the approach in \cite{yao2021wenet} to achieve faster and improved convergence. Our encoder backbone employs the large version, consisting of 17 encoder layers, 8 attention heads, and 512 encoder dimensions.  Our training strategy incorporates a hybrid loss function, combining both CTC and AED losses, with a balancing hyperparameter $\lambda = 0.3$:
\begin{align}
L_{total} &= \lambda L_{CTC} + (1 - \lambda) L_{AED}
\end{align}
During inference, we exclusively employ the CTC decoder, and performance evaluation is conducted using greedy search. \\
\textbf{Training:} 
\begin{table*}[ht]
\vspace{-23pt}
\centering
\caption{\label{tab:allresults} WER results on the experimental test sets. A fixed 20-second buffer and a 6-minute audio segment are specifically used for decoding in full-context models. Values in square braces represent $[l_{att}, c, r]$.}
\vspace{-5pt}
    \begin{tabular}{ l c c c c c c c c c c  l}
    \hline
     \multirow{2}{*}{\textbf{Model}}  & \textbf{test}  & \textbf{test} & \textbf{tedlium} & \textbf{earnings} & \textbf{earnings} & \multirow{2}{*}{\textbf{reading}} & \textbf{conver-} & \textbf{tele-} & \multirow{2}{*}{\textbf{youtube}} & \multirow{2}{*}{\textbf{average}} \\ 
     & \multicolumn{1}{c}{\textbf{clean}} &  \multicolumn{1}{c}{\textbf{other}} &  \multicolumn{1}{c}{\textbf{v3}} & \multicolumn{1}{c}{\textbf{21}} & \multicolumn{1}{c}{\textbf{22}} &  & \textbf{sation} & \textbf{phony}  &   \\
    \hline
    \multirow{1}{*}{\textbf{Conformer} \cite{49414}}  & 2.77 & 6.93 & 24.03 & 39.39 & 54.91 & \textbf{5.00} & 7.25 & 12.34 & 16.89 & 18.83 \\
    \multirow{1}{*}{\textbf{Efficient Conformer} \cite{burchi2021efficient}}  & 2.71 & 6.95 & 23.40 & 36.64 & 52.53 & 5.25 & 7.35 & \textbf{11.63}  & 16.19 & 18.07 \\
    \multirow{1}{*}{\textbf{Squeezeformer} \cite{kim2022squeezeformer}}  & 2.87 & 7.17 & 23.50 & 38.09 & 53.44 & 5.20 & 7.29 & 11.72 & 16.83 & 18.46 \\
    \hline
    \textbf{ChunkFormer} 
     & \textbf{2.68} & \textbf{6.88} & 23.37 & 38.67 & 54.55 & 5.08 & 7.19 & 11.86 & 15.00 & 18.36 \\
    \text{\hspace{0.3cm}$[128, 256, 128]$} & 2.74 & 6.90 & 21.38 & 32.04 & 48.68 & 5.11 & 7.19 & 11.85 & 14.20 & 16.68 \\
    \text{\hspace{0.3cm}$[256, 128, 128]$} & 2.74 & 6.90 & 21.41 & 32.05 & 48.67 & 5.13 & 7.19 & 11.84 & 14.22 & 16.68 \\
    \text{\hspace{0.3cm}$[128, 64, 128]$} & 2.74 & 6.90 & \textbf{21.32} & \textbf{31.73} & \textbf{48.56} & 5.05 & \textbf{7.17} & 11.85 & \textbf{14.05} & \textbf{16.60} \\


    \hline
    \end{tabular}
\vspace{-14pt}

\end{table*}
Our audio front end employs 80-dimensional filter-bank features with 25ms FFT windows and a 10ms frameshift, enhanced by Speed-Perturb and SpecAugment techniques. The Byte-Pair Encoding vocabulary includes 5000 tokens, each represented by a 256-dimensional embedding. 
We train full-context models from scratch using the WeNet 2.0 toolkit \cite{zhang2022wenet} on 8 NVIDIA H100 GPUs with mixed precision. Training lasts 200 epochs and 400,000 steps for small and large datasets, respectively utilizing the Adam optimizer and Noam warm-up scheduler. The learning rate reaches a peak of 1e-3 after 15,000 steps (small) and 25,000 steps (large). The masked batch limited context models are fine-tuned from pre-trained full context models for 50 epochs (small) and 100,000 steps (large) with a peak learning rate reset to 1e-5. We also apply this step to the full context model to verify the improvements aren't solely due to extended training. The final model is obtained by averaging the last 50 (small) and 10 (large) checkpoints.
Dynamic context training is used to improve the robustness of limited context models by adjusting $l_{att}$, $c$, and $r$ during training.
\subsection{Results}
\label{sec:results}

\begin{table}[t]
\centering
\caption{\label{tab:duration} Maximum audio duration (in minutes) which can be processed by the model encoder with batch size 1 on an 80GB. Values in square braces represent $[l_{att}, c, r]$.}
\vspace{-4pt}
    \begin{tabular}{ l  c c}
    
    \multirow{1}{*}{\textbf{Model}} & \multirow{1}{*}{\textbf{Params}}  & \textbf{Max duration}\\
    \hline
    \multirow{1}{*}{\textbf{Conformer} \cite{49414}} 
    & \multirow{1}{*}{115M} & 15 \\
    \hline
    \multirow{1}{*}{\textbf{FastConformer} \cite{rekesh2023fast}} & \multirow{2}{*}{109M} & 25  \\
    \text{\hspace{0.3cm}$[128, 1, 128]$} &  & 675 \\
    \hline
    \multirow{1}{*}{\textbf{ChunkFormer}} & \multirow{3}{*}{110M}
    & 25 \\
    \text{\hspace{0.3cm}$[256, 128, 128]$}  & & \textbf{760}  \\
    \text{\hspace{0.3cm}$[128, 64, 128]$}  & & \textbf{980}  \\
    \hline    
    \end{tabular}
\vspace{-13pt}

\end{table}
The results in Table \ref{tab:allresults} show that ChunkFormer consistently outperforms recent state-of-the-art models in both long-form and short-medium-form speech recognition tasks, across low-resource and large-scale datasets. In the full-context setting, ChunkFormer achieves comparable performance to Conformer, Efficient Conformer, and Squeezeformer on most test datasets, with an average WER of 18.36\%. However, when switching to a limited chunk context for convolution and attention, ChunkFormer exhibits a more pronounced advantage.  With the context configuration of $[128, 64, 128]$, ChunkFormer achieves the lowest WER on the Tedlium-v3, Earnings-21, and Earnings-22 datasets, recording 21.32\%, 31.73\%, and 48.56\% respectively. This significantly outperforms other models, with a notable 7.7\% absolute reduction in WER on the Earnings-21 dataset compared to the baseline Conformer model. The improvements in long-form tasks stem from ChunkFormer's limited-context and streaming-style decoding, which preserves relevant context between chunks and aligns with its training, eliminating the need for buffering or fixed-length segmentation.
In large-scale training, ChunkFormer also maintains its competitiveness in both short-medium and long-form tasks, with WERs of 5.05\%, 7.17\%, 11.85\%, and 14.05\% on the Reading, Conversation, Telephony, and YouTube datasets, respectively. These results highlight ChunkFormer’s ability to enhance long-form transcription performance while maintaining high accuracy across shorter tasks, validating its robustness and efficiency across diverse speech recognition scenarios.
We also observe that larger chunk sizes yield slightly worse results than smaller ones. Since chunk sizes of 128 or 256 correspond to approximately 10 and 20 seconds of audio, respectively. Any audio shorter than chunk size will be processed as a single chunk, making it no different from full-context learning. Consequently, ChunkFormer cannot effectively learn the masked batch chunk-context patterns from such shorter audios.
\begin{table}[h]
\centering

    \caption{\label{tab:maskedbatch} Memory usage (GB), execution time (second), and FLOPS (T) comparisons between naive batch FastConformer, naive batch ChunkFormer, and masked batch ChunkFormer for processing a batch of 1s, 30s, 1m, 15m, 30m, and 1h audios. ChunkFormer utilizes the context size of $[128, 64, 128]$.}
    
\vspace{-4pt}

\resizebox{\columnwidth}{!}{
    \begin{tabular}{ l  c  c  c  c  c  c}
    \multicolumn{1}{c}{\multirow{2}{*}{\textbf{Model}}} & \multicolumn{3}{c}{\textbf{Full Model}} & 
    \multicolumn{3}{c}{\textbf{Encoder only}} \\
    \cline{2-7}
    & \textbf{mem} & \multirow{1}{*}{\textbf{time}} & \multirow{1}{*}{\textbf{flops}} & \textbf{mem} & \multirow{1}{*}{\textbf{time}} & \multirow{1}{*}{\textbf{flops}} \\
    \hline
    \multirow{1}{*}{\textbf{FastConformer}\cite{rekesh2023fast}}
    & 73.4 & 2.4 & 64.3 & 26.4 & 2.1 & 61.5 \\
    \hline
    \multirow{1}{*}{\textbf{ChunkFormer}} & 34.5 & 2.7 & 65.2 & 25.8 & 2.2 & 56.7 \\
    \text{\hspace{0.3cm}+ Masked batch} & \textbf{19.6} &  \textbf{0.8} & \textbf{19.3} & \textbf{8.1} & \textbf{0.7} & \textbf{16.8} \\
    \hline
    \end{tabular}
}
\vspace{-13pt}
\end{table}

The results from Table \ref{tab:duration} demonstrate that ChunkFormer significantly outperforms both Conformer and FastConformer in terms of maximum audio duration processed. With 110M parameters, ChunkFormer achieves a maximum duration of 980 minutes using the $[128, 64, 128]$ context configuration, compared to 15 minutes with Conformer and 675 minutes with FastConformer. This highlights ChunkFormer's efficiency and capability in handling longer audio sequences within the same hardware constraints. The superior performance of ChunkFormer in processing longer audio durations, despite its larger context size in attention layers, stems from its chunk-wise mechanism. In contrast, the token-wise processing in FastConformer is less efficient for parallel processing. 

Additionally, we experimented with comparing the efficiency of the naive batch method and our proposed masked batch method, as well as their compatibility with the convolution-splitting technique described in Section 2.4. The results, presented in Table \ref{tab:maskedbatch}, demonstrate the significant advantages of the masked batch ChunkFormer model. The naive batch method, due to its padding requirements, leads to 1.5 to 3.7 times higher memory consumption, 3 to 3.4 times longer execution times, and 1.8 to 3.7 times greater FLOPs when processing a batch of 1s, 30s, 1m, 15m, 30m, and 1h audios, compared to the masked batch method. Notably, the convolution channel splitting in FastConformer demands significantly more memory compared to the batch splitting in ChunkFormer. Given the potential for unfair memory comparisons caused by different convolution-splitting methods, we excluded the downsampling module and focused solely on the encoder's efficiency. Under these conditions, the naive Batch FastConformer and ChunkFormer exhibit similar results in memory usage and execution time, but ChunkFormer achieves lower FLOPs, underscoring the efficiency of chunk-wise over token-wise processing. Moreover, the masked batch method maintains its superior efficiency in both memory usage, execution time, and FLOPS for both full model and encoder-only settings, establishing it as the most optimized model among the three evaluated.
\section{Conclusions}
In summary, we presented ChunkFormer, an efficient ASR model designed for long-form transcription on resource-constrained hardware. By utilizing chunk-based processing with relative right context and introducing the masked batch decoding technique, ChunkFormer significantly reduces memory consumption and execution time, offering a 3.7x improvement over existing models. Experimental results show significant performance gains in WER and efficiency, processing up to 980 minutes of audio on a single GPU, far surpassing current state-of-the-art models. ChunkFormer's scalability and efficiency make it highly applicable for large-scale ASR deployments, offering a cost-effective solution for industrial use cases.

\bibliographystyle{IEEEbib}
\bibliography{refs}
\nocite{*}

\end{document}